\documentstyle[preprint,aps,psfig]{revtex}

\begin{document}
\draft
\title{Nuclear Matter on a Lattice}

\author{H.-M. M\"uller${}^1$, S. E. Koonin${}^1$, R. Seki${}^{1,2}$, and U. van Kolck${}^1$}

\address{${}^1$ W. K. Kellogg Radiation Laboratory,\\
California Institute of Technology,\\ Pasadena, CA 91125, USA \\
${}^2$ Department of Physics and Astronomy,\\ California State
University, Northridge,\\ Northridge, CA 91330, USA}

\date{\today}

\maketitle

\normalsize

\begin{abstract}
We investigate nuclear matter on a cubic lattice. An exact thermal
formalism is applied to nucleons with a Hamiltonian that
accommodates on-site and next-neighbor parts of the central, spin-
and isospin-exchange interactions. We describe the nuclear matter
Monte Carlo methods which contain elements from shell model Monte
Carlo methods and from numerical simulations of the Hubbard model.
We show that energy and basic saturation properties of nuclear
matter can be reproduced. Evidence of a first-order phase
transition from an uncorrelated Fermi gas to a clustered system is
observed by computing mechanical and thermodynamical quantities
such as compressibility, heat capacity, entropy and grand
potential. We compare symmetry energy and first sound velocities
with literature and find reasonable agreement.
\end{abstract}

\narrowtext

\section{Introduction}

Properties of nuclear matter have been deduced from different
approaches. The volume term of the semi-empirical mass formula
\cite{weizs35,bethe36} predicts a binding energy of $16 \; {\rm
MeV}$, and calculations of finite nuclei estimate the equilibrium
density to be  $\rho_0 = 0.16 \; {\rm fm^{-3}}$. However,
properties of finite nuclei  are strongly influenced by finite
size effects like the surface effect, and it is therefore
difficult to estimate energies and saturation density of nuclear
matter from nuclei. Many-body calculations of nuclear matter are
based on sophisticated potentials (such as the Argonne AV14 and
AV18 and Urbana UV14 potentials) and use Bethe-Brueckner-Goldstone
theory \cite{brueckner,bg57,goldstone} and hypernetted chain
approximations \cite{wff,aakmal} to calculate {\em ground state}
properties. Lattice gas calculations \cite{ttskuo,campi,pan4}
attempt a {\em thermal} description of nuclear matter. They work
with much simpler Hamiltonians, incorporating isospin-1 or
Hubbard-like interactions. These calculations aim at the
investigation of a liquid-gas phase transition of nuclear matter
expected to take place at subnuclear densities and low
temperatures. They are {\em classical}, not quantum mechanical,
putting in kinetic terms by hand or sampling them from a
Maxwell-Boltzmann distribution. These types of calculations use
Monte-Carlo-like algorithms and show that the inclusion of a
kinetic term is crucial to observe a phase transition.

This paper describes first results of a calculation of infinite
nuclear matter that combines both the usage of a more realistic
Hamiltonian and the exact, thermal treatment of the many-body
problem on a lattice. In the last few years, the shell model Monte
Carlo (SMMC) method has been successfully developed
\cite{cwjohnson,gladys,weormand,yalhassid,physrep} to give a
powerful alternative to direct diagonalization procedures which
suffer from the fact that the many-body space scales so
unfavorably with the number of single-body states considered.
Direct diagonalization methods can only address very light nuclei
or nuclei with a closed shell and only a few valence nucleons. The
SMMC avoids this combinatorial scaling (in storage and computation
time) and makes it possible to investigate structural properties
of nuclei far beyond the few-nucleon system. The SMMC enforces the
Pauli-principle exactly, and concentrates on the evaluation of
thermal averages of observables. This would be the main purpose of
a nuclear matter investigation too: Not focusing on obtaining a
wave function, a thermal formalism is useful for a study of
nuclear matter, because the equation of state is of main interest,
which clearly depends on density and temperature. Moreover, the
consideration of a large piece of infinite nuclear matter in {\em
coordinate} space reduces finite size effects that appear after
imposing periodic boundary conditions. A formalism written in
momentum space has the disadvantage that two- or many-body
correlations cannot be calculated directly: Clustering (and
therefore a possible liquid-gas transition) is not as easily
calculated and observed as in the coordinate space representation.

The following concept is pursued for our nuclear matter
calculation: The quantum mechanical and exact treatment with the
full Hamiltonian, kinetic and potential term, should be a
prerequisite for a successful description of the physical system.
In a coordinate representation nucleons shall interact with a
potential that eventually comes as close to a realistic nuclear
interaction (like AV18) as possible. The partition function along
with observables of interest shall be calculated in the grand
canonical ensemble, in order to control temperature $T$ and
density $\rho$. The latter is to be adjusted on average via the
chemical potential $\mu$. The many-body problem shall be solved
exactly using Monte Carlo methods similar to those used in the
SMMC applications. At the same time, realizing that the emerging
equations eventually have to be solved on a computer, one should
take into account that space will be discretized, and advantage
should be taken of the available technology that has been employed
for the Hubbard and other models in condensed matter physics. This
paper is to be viewed as a first step of a full thermal
description of nuclear matter in which we constrain our potential
parameters to a reasonable shape of the energy as a function of
density, including the correct saturation point.

\section{Theory of Nucleonic Matter on a Lattice}
The general concept of the nuclear matter calculation consists of
nucleons interacting via a variety of components of the nuclear
two-body potential. While it should be the ultimate goal to use a
potential that fits the nucleon-nucleon scattering data best
\cite{wss}, at the first stage we concentrate on few parts of the
interactions, namely central, spin- and isospin-exchange. The
degrees of freedom of the nucleon are its spin, isospin as well as
the spatial coordinate.

Subnuclear degrees of freedom are not explicitly incorporated. The
lightest meson, the pion, facilitates an interaction with a range
of $r \approx 1.4 \;{\rm fm}$ which is of same order as the
lattice spacing of the applications in this paper. Since the
system is ultimately regularized on a lattice, the argument can be
made that all subnuclear degrees of freedom are integrated out,
resulting in a strong on-site and weaker next-neighbor
interaction. The lattice spacing, here an additional fitting
parameter like the potential parameters, is chosen to be of $a =
1.842 \; {\rm fm}$. This particular lattice spacing sets the
half-filling of the lattice at $\rho=2\rho_0=0.32 \; {\rm
fm^{-3}}$. Other settings of fillings have been tried, but turned
out to reproduce the saturation curve less well.

In this section we specify the Hamiltonian of the system and
describe the nuclear matter Monte Carlo method (called NMMC
hereafter), which consists of the thermal formalism to express the
grand canonical partition function as an integral over single-body
evolution operators. At its center stands the
Hubbard-Stratonovitch transformation, which is used to reduce the
many-body problem to an effective one-body problem. The details of
the Monte Carlo procedure, which is used to evaluate the resulting
multi-dimensional integral, are explained.

\subsection{Hamiltonian}
\label{hamilsec} We consider a three-dimensional cubic lattice of
spacing $a$ and assume periodic boundary conditions, which result
in a three-dimensional toroidal configuration. The coordinate
$\vec{x}$ and the momentum $\vec{p}$ are discretized as
\begin{equation}
\label{discrspc} \vec{x} \rightarrow a \vec{m} \equiv \vec{x}_m,
\end{equation}
\begin{equation}
\label{discrmom} \vec{p} \rightarrow  \left( \frac{2\pi}{Na}
\right) \vec{k} \equiv \vec{p}_k,
\end{equation}
such that
\begin{equation}
\vec{x} \cdot \vec{p} = \frac{2\pi}{N} \times {\rm integer},
\end{equation}
where $N$ is the number of lattice points in each spatial
direction, and $\vec{m}$ and $\vec{k}$ are vectors with integer
components.

The nucleons have mass $m_N$, spin $\sigma = \pm \frac{1}{2}$ and
isospin $\tau = \pm \frac{1}{2}$ . The Hamiltonian,
\begin{equation}
\hat{\cal H} = \hat{\cal K} + \hat{\cal V},
\end{equation}
is expressed in second quantization and contains kinetic and
potential operators. The kinetic term is written as
\begin{equation}
\hat{\cal K} = - \frac{\hbar^2}{2m_N} \sum_{\sigma \tau} \int {\rm
d}\vec{x} \; \psi^\dagger_{\sigma \tau}(\vec{x}) \vec{\nabla}^2
\psi_{\sigma \tau}(\vec{x}).
\end{equation}
The fermion operator $\psi^\dagger_{\sigma \tau}(\vec{x})$ creates
a nucleon of spin and isospin $(\sigma, \tau)$ at location
$\vec{x}$, while its adjoint $\psi_{\sigma \tau}(\vec{x})$
destroys it. This equation is discretized on the lattice by the
symmetric 3-point formula for the second derivative, and the
integral is replaced by a finite sum, which results in
\begin{equation}
\hat{\cal K} = - t_0 \sum_{\sigma \tau} a^3 \sum_{\vec{x}_n,i=1
\cdots 3} \psi^\dagger_{\sigma \tau}(\vec{x}_n) \left[
\psi_{\sigma \tau}(\vec{x}_n+a \vec{e}_i) - 2 \psi_{\sigma
\tau}(\vec{x}_n) + \psi_{\sigma \tau}(\vec{x}_n-a \vec{e}_i)
\right]
\end{equation}
with
\begin{equation}
t_0 = \frac{\hbar^2}{2m_Na^2}.
\end{equation}
Here, the orthogonal unit vectors $\{\vec{e}_i\}$ span the
three-dimensional space.

While the form of the nuclear potential is generally given, we
here are limited by current computational constraints. The
treatment of a full Hamiltonian, as it is represented in the
Argonne potential, for example, is computationally impossible with
currently available computer power, but may be feasible in a few
years. We chose
\begin{equation} \hat{\cal V} = \hat{\cal V}_c +
\hat{\cal V}_\sigma.
\end{equation}
The first part is the central potential ($\hat{\cal V}_c$),
followed by the spin-exchange ($\hat{\cal V}_\sigma$). The general
form for the scalar potential,
\begin{equation}
\hat{\cal V}_c = \frac{1}{2} \sum_{\sigma \tau \sigma^\prime
\tau^\prime} \int {\rm d}\vec{x} \; \int {\rm d}\vec{x}^\prime \;
\psi^\dagger_{\sigma \tau}(\vec{x}) \psi^\dagger_{\sigma^\prime
\tau^\prime}(\vec{x}^\prime) V_c(\vec{x}-\vec{x}^\prime)
\psi_{\sigma^\prime \tau^\prime}(\vec{x}^\prime)\psi_{\sigma
\tau}(\vec{x}),
\end{equation}
can be written in terms of the density
\begin{equation}
\hat{\rho}(\vec{x}) = \sum_{\sigma \tau} \hat{\rho}_{\sigma
\tau}(\vec{x}) = \sum_{\sigma \tau} \psi^\dagger_{\sigma
\tau}(\vec{x})\psi_{\sigma \tau}(\vec{x}).
\end{equation}
The purpose of doing so is to cast the potential in linear and
quadratic terms, as the Hubbard-Stratonovitch transformation can
only be performed on quadratic terms. Using the fermion
anticommutation relation, the potential then becomes
\begin{equation}
\hat{\cal V}_c = \frac{1}{2} \int {\rm d}\vec{x} \; \int {\rm
d}\vec{x}^\prime \; V_c(\vec{x}-\vec{x}^\prime)
\hat{\rho}(\vec{x}) \hat{\rho}(\vec{x}^\prime)  - \frac{1}{2} \int
{\rm d}\vec{x} \; V_c(0) \hat{\rho}(\vec{x}).
\end{equation}
The last term is the self-energy and is a consequence of the Pauli
principle. The discretized version of this equation is
\begin{equation}
\label{potentialoperator} \hat{\cal V}_c = \frac{a^6}{2}
\sum_{\vec{x}_n, \vec{x}^\prime_n} V_c(\vec{x}_n-\vec{x}^\prime_n)
\hat{\rho}(\vec{x}_n) \hat{\rho}(\vec{x}^\prime_n)  -
\frac{a^3}{2} \sum_{\vec{x}_n} V_c(0) \hat{\rho}(\vec{x}_n).
\end{equation}
We assume a Skyrme-like on-site and next-neighbor interaction
\begin{equation}
\label{skyrmelike} V_c\left(\vec{x}_n-\vec{x}^\prime_n\right) =
V^{(0)}_c \delta \left(\vec{x}_n-\vec{x}^\prime_n\right) +
V^{(2)}_c \left(\nabla^2_{\vec{x}_n} \delta
\left(\vec{x}_n-\vec{x}^\prime_n\right)\right),
\end{equation}
whose discretized form is
\begin{equation}
\label{skyrme} V_c(\vec{x}_n-\vec{x}^\prime_n) =
\frac{V^{(0)}_c}{a^3} \delta_{\vec{x}_n,\vec{x}^\prime_n} +
\frac{V^{(2)}_c}{a^5} \sum_{i=1}^3 \left\{
\delta_{\vec{x}_n+a\vec{e}_i,\vec{x}^\prime_n} -
2\delta_{\vec{x}_n,\vec{x}^\prime_n} +
\delta_{\vec{x}_n-a\vec{e}_i,\vec{x}^\prime_n} \right\}.
\end{equation}
The parentheses in Eq.~(\ref{skyrmelike}) indicate that the
Laplace operator only acts on the $\delta$-function, but not on
any following parts. Inserting equation (\ref{skyrme}) into
(\ref{potentialoperator}) gives
\begin{eqnarray}
\label{vcfin} \hat{\cal V}_c =  \frac{V^{(0)}_c}{2}
\sum_{\vec{x}_n} a^3 \hat{\rho}(\vec{x}_n)^2 &-& \frac{V^{(2)}_c a
}{2} \sum_{\vec{x}_n}  \sum_{i=1}^3
\left(\hat{\rho}\left(\vec{x}_n+a\vec{e}_i\right)-
\hat{\rho}\left(\vec{x}_n\right)\right)^2 \nonumber\\ &-&
\frac{1}{2} \left( V^{(0)}_c - 6\frac{V^{(2)}_c}{a^2} \right)
\sum_{\vec{x}_n} \hat{\rho}(\vec{x}_n),
\end{eqnarray}
Here, we applied periodic boundary conditions.

The spin-exchange part of the potential is handled in a very
similar way. Starting from
\begin{equation}
\label{sep} \hat{\cal V}_\sigma =  \frac{1}{2} \sum_{{\xi \tau
\xi^\prime \tau^\prime} \atop {\kappa \lambda \kappa^\prime
\lambda^\prime}} \int {\rm d} \vec{x} \; \int {\rm
d}\vec{x}^\prime
 \; \psi^\dagger_{\xi \tau}(\vec{x})
\psi^\dagger_{\xi^\prime \tau^\prime}(\vec{x}^\prime)
V_\sigma(\vec{x}-\vec{x}^\prime) \vec{\bf \sigma}_{\xi \tau \kappa
\lambda} \cdot \vec{\bf \sigma}_{\xi^\prime \tau^\prime
\kappa^\prime \lambda^\prime} \psi_{\kappa^\prime
\lambda^\prime}(\vec{x}^\prime)\psi_{\kappa \lambda}(\vec{x}),
\end{equation}
we write the potential in form of spin densities
\begin{equation}
\hat{\rho}^{(\alpha)}_\sigma(\vec{x}) = \sum_{\xi \tau \kappa
\lambda} \psi^\dagger_{\xi \tau}(\vec{x}) {\bf
\sigma}^{(\alpha)}_{\xi \tau \kappa \lambda} \psi_{\kappa
\lambda}(\vec{x}), \; \alpha = 0,+,-,
\end{equation}
where ${\bf \sigma}^{(\alpha)}_{\xi \tau \kappa \lambda}$ are the
elements of a generalized Pauli spin-isospin matrix, which acts on
a 4-vector representing all spin/isospin states of a nucleon. We
assume the the same spatial dependence (\ref{skyrmelike}) as for
the central part, and finally arrive at
\begin{eqnarray}
\label{vsfin} \hat{\cal V}_\sigma &=&  \frac{V^{(0)}_\sigma}{2}
\sum_{\vec{x}_n} a^3 \left(
\hat{\rho}^{(0)}_\sigma\left(\vec{x}_n\right)^2 + 2
\left(\hat{\rho}^{(+)}_\sigma\left(\vec{x}_n\right) +
\hat{\rho}^{(-)}_\sigma\left(\vec{x}_n\right) \right)^2 \right)
\nonumber\\ &-& \frac{V^{(2)}_\sigma a}{2} \sum_{\vec{x}_n}
\sum_{i=1}^{3} \left( \left( \hat{\rho}^{(0)}_\sigma(\vec{x}_n + a
\vec{e}_i) - \hat{\rho}^{(0)}_\sigma(\vec{x}_n) \right)^2 \right.
\nonumber\\ &+& \left. 2 \left(
\hat{\rho}^{(+)}_\sigma\left(\vec{x}_n + a \vec{e}_i \right) +
\hat{\rho}^{(-)}_\sigma\left(\vec{x}_n + a \vec{e}_i \right) -
\hat{\rho}^{(+)}_\sigma\left(\vec{x}_n \right) -
\hat{\rho}^{(-)}_\sigma\left(\vec{x}_n \right) \right)^2 \right)
\nonumber\\ &-& \frac{3}{2} \left( V^{(0)}_\sigma -
6\frac{V^{(2)}_\sigma}{a^2} \right) \sum_{\vec{x}_n}
\hat{\rho}\left(\vec{x}_n\right).
\end{eqnarray}
Other components of the potential can be included in a similar
way.

\subsection{Nuclear Matter Monte Carlo Method}
In order to study thermal properties of nuclear matter, the grand
canonical partition function at a given temperature $T=\beta^{-1}$
needs to be determined,
\begin{equation}
Z = {\rm \hat{Tr}} \left[ \exp \left( -\beta \left( \hat{\cal H} -
\sum_{\sigma \tau} \mu_\tau \hat{\cal N}_{\sigma \tau} \right)
\right) \right] \equiv {\rm \hat{Tr}} \left[ \hat{U} \right],
\end{equation}
with $\hat{\cal N}_{\sigma \tau} = \sum_{\vec{x}_n}
\psi^\dagger_{\sigma \tau}(\vec{x}_n)\psi_{\sigma
\tau}(\vec{x}_n)$ and $\mu_\tau$ as the isospin-dependent chemical
potential. $\hat{U}$ is called the imaginary-time evolution
operator of the system and is a many-body operator. In the present
study the Hamiltonian $\hat{\cal H}$ contains one- and two-body
operators as described Section~\ref{hamilsec}, and the trace is
taken over all many-body states as indicated by a caret. The
partition function $Z$ is an exponential over all one- and
two-body operators (and therefore interactions) present in the
system. It is impossible to deal with the partition function $Z$
in this form, because the number of many-body correlations that
have to be kept track of grows rapidly with system size. We
therefore seek an expression for $Z$ that is based on a
single-particle representation, and we will replace the many-body
problem with that of non-interacting nucleons that are coupled to
a heat bath of auxiliary fields. This involves rewriting $Z$ as a
multi-dimensional integral.

We start by dividing the evolution operator into $n_t$ time
slices:
\begin{equation}
\hat{U} = \exp \left( -\beta \left(\hat{\cal H} - \sum_{\sigma,
\tau} \mu_\tau \hat{\cal N}_{\sigma \tau} \right) \right) = \left[
\exp \left(- \Delta \beta \left( \hat{\cal H} - \sum_{\sigma,
\tau} \mu_\tau \hat{\cal N}_{\sigma \tau} \right) \right)
\right]^{n_t},
\end{equation}
with $\Delta \beta n_t = \beta$. The Trotter approximation
\cite{trotter,suzuki} is used to separate one-body (kinetic energy
and chemical potential) in  $\hat{\cal S} \equiv \hat{\cal K} -
\sum_{\sigma, \tau} \mu_\tau \hat{\cal N}_{\sigma \tau}$ and
two-body terms (potential) in $\hat{\cal H}$:
\begin{equation}
\label{trotter} \exp \left(- \Delta \beta \left( \hat{\cal H} -
\sum_{\sigma, \tau} \mu_\tau \hat{\cal N}_{\sigma \tau}\right)
\right) = \exp \left( -\Delta \beta \left(\hat{\cal S}+\hat{\cal
V}\right)\right) \approx \exp \left(-\Delta \beta \hat{\cal
S}\right)\exp \left(-\Delta \beta \hat{\cal V}\right).
\end{equation}
Equation (\ref{trotter}) is valid to order $\Delta \beta$, but
becomes exact in the limit $\Delta \beta \rightarrow 0$.

The propagator of each time slice for the potential, $\exp
(-\Delta \beta \hat{\cal V})$, is manipulated by applying the
Hubbard-Stratonovitch (HS) transformation
\cite{hubbard,stratonovitch} on each term, replacing it with a
multi-dimensional integral over a set of auxiliary fields.

As an example, we describe the transformation by taking the
on-site part of $\hat{\cal V}_c$ at one particular site
$\vec{x}_m$. Using $\alpha \equiv \Delta \beta V^{(0)}_c/2$ and
defining
\begin{equation}
\label{salpha} S_{\alpha} = \left\{ \begin{array}{ll} \pm i & {\rm
if} \; \alpha
> 0 \\ \pm 1 & {\rm if} \; \alpha < 0, \end{array} \right.
\end{equation}
the propagator for this single interaction is written as
\begin{eqnarray}
\label{uvc0} \Delta \hat{U} \left(\vec{x}_m\right) & \equiv & \exp
\left( -\Delta \beta \frac{V^{(0)}_c}{2} \hat{\rho}^2
\left(\vec{x}_m\right) \right) =  \sqrt{\frac{| \alpha |}{\pi}}
\int_{-\infty}^\infty {\rm d} \chi \exp \left( - \alpha
\hat{\rho}^2 \left(\vec{x}_m\right) - | \alpha | \left( \chi +
S_{\alpha} \hat{\rho}\left(\vec{x}_m\right) \right)^2 \right)
\nonumber
\\ & = & \sqrt{\frac{| \alpha |}{\pi}}
\int_{-\infty}^\infty {\rm d} \chi  \exp \left( - | \alpha |
\left[ \chi^2 + 2 S_{\alpha} \chi \hat{\rho}
\left(\vec{x}_m\right) \right] \right).
\end{eqnarray}
The last line of Eq.~(\ref{uvc0}) reveals that the evolution
operator is now expressed in terms of the exponential of a
one-body operator and an integration over the auxiliary field
$\chi$. It has become a one-body propagator that corresponds to
non-interacting nucleons coupled to this field. Since the integral
is calculated with Monte Carlo methods, the field fluctuates
according to a weight that is to be specified, hence the picture
of a heat bath.

It has to be emphasized that $\hat{\rho}\left(\vec{x}_m\right)$
here represents a one-body operator for a subset ($\hat{\cal V}_c$
in this case) of the full interaction. Each quadratic term in
(\ref{vcfin}) and (\ref{vsfin}) has to be replaced by an integral.
At a given lattice site $\vec{x}_m$, there are twelve auxiliary
fields to form the full interaction, four for $\hat{\cal V}_c$ and
eight for $\hat{\cal V}_\sigma$. Nucleons are now coupled to a big
ensemble of auxiliary fields through which the interaction of the
nucleons is mediated. The $\Delta \hat{U}$'s are then multiplied
together to form the evolution operator for one time slice
$\hat{U}\left(\beta_m \right)$ at $\beta_m = m\Delta \beta$, which
is expressed only in terms of single-body matrices, and
ultimately, all time slices are multiplied together to form
$\hat{U}$:
\begin{equation}
\label{totalu} \hat{U} = \left[ \exp \left(- \Delta \beta
\hat{\cal H} \right) \right]^{n_t} = \int {\cal D} \left[ \chi
\right] G\left(\chi\right) \hat{U}_\chi \left(\beta, 0\right)
\end{equation}
with the integration measure
\begin{equation}
\label{intmeas} {\cal D} \left[ \chi \right] =  \prod_{m=1}^{n_t}
\prod_{\vec{x}_n} \prod_{i} {\rm d\chi}_{m, \vec{x}_n, i}
\sqrt{\frac{|\alpha_i|}{\pi}}.
\end{equation}
The $\alpha_i = \Delta \beta V_i/2,\; V_i \in \left[V^{(0)}_c,
V^{(2)}_c, V^{(0)}_\sigma, \cdots \right]$, are the
interaction-specific coupling strengths of auxiliary fields to
nucleons, and the index $i$ enumerates all fields at a particular
site. The Gaussian factor $G$ is given by
\begin{equation}
\label{gaussfac} G\left(\chi\right) = \prod_{m=1}^{n_t}
\prod_{\vec{x}_n} \prod_{i} \exp \left( - |\alpha_i| \chi^2_{m,
\vec{x}_n, i} \right),
\end{equation}
and the one-body propagator is
\begin{equation}
\hat{U}_\chi \left(\beta,0\right) = \hat{U}
\left(\beta_{n_t}\right)\hat{U}\left(\beta_{n_t-1}\right) \cdots
\hat{U}\left(\beta_1\right).
\end{equation}
Note that Eq.~(\ref{totalu}) only becomes exact in the limit of an
infinite number of time slices, $n_t \rightarrow \infty$. For a
finite $n_t$, the Hubbard-Stratonovitch approximation is valid
only to order $\Delta \beta$.

In the practical implementation of Eq.~(\ref{uvc0}), a discrete
Hubbard-Stratonovitch transformation is used instead of the
continuous form because it turns out to use much fewer
de-correlation sweeps, as explained below.

A thermal observable $\langle \hat{O} \rangle$ is expressed as
\cite{gladys,gubernatis}
\begin{equation}
\label{therobs} \langle \hat{O}\rangle = \frac{1}{Z} {\rm
\hat{Tr}} \left[\hat{O} \exp \left( -\beta \left(\hat{{\cal H}}-
\sum_{\sigma, \tau} \mu_\tau \hat{\cal N}_{\sigma \tau}\right)
\right)\right] = \frac {\int {\cal D} [ \chi ] G(\chi) \langle
\hat{O}(\chi) \rangle \xi(\chi)}{\int {\cal D} [ \chi ]
G(\chi)\xi(\chi)}
\end{equation}
and has the integration measure of Eq.~(\ref{intmeas}) and
Gaussian factor of Eq.~(\ref{gaussfac}). The expectation value of
any operator in second quantization can be obtained with the help
of Wick's theorem, and the resulting one-body densities are
\cite{gladys}
\begin{equation}
\langle \psi^\dagger_{\sigma \tau} \left(\vec{x}_n \right)
\psi_{\sigma^\prime \tau^\prime} \left(\vec{x}_m\right)
\rangle_{\chi} = \left\{ \left[ {\bf 1} + {\bf
U}_{\chi}\left(\beta,0\right) \right]^{-1} {\bf U}_{\chi} \left(
\beta,0 \right) \right\}_{(\sigma^\prime \tau^\prime,
\vec{x}_m),(\sigma \tau, \vec{x}_n)}.
\end{equation}
The bold face ${\bf U}_\chi\left(\beta,0\right)$ is the
single-body matrix representation of
$\hat{U}_\chi\left(\beta,0\right)$. Observables of the system are
chosen to be the number of neutrons and protons and all components
of the energy.

The integrals in (\ref{therobs}) are evaluated using the
Metropolis algorithm \cite{metropolis}. The basic idea involves
sampling the integrand of (\ref{therobs}),
\begin{equation}
\langle \hat{O}(\chi) \rangle = \frac{ \hat{\rm Tr}
\left[\hat{O}\hat{U}_\chi\left(\beta,0\right)\right]}{\hat{\rm Tr}
\left[\hat{U}_\chi\left(\beta,0\right)\right]},
\end{equation}
within the boundaries of the integration volume according to a
positive-definite weight
\begin{equation}
W\left(\chi\right) = \left\{ \begin{array}{ll} | G(\chi) \xi(\chi)
| & {\rm for \; continuous \; HS}\\ | \xi(\chi) |& {\rm for \;
discrete \; HS},
\end{array}
\right.
\end{equation}
with
\begin{equation}
\label{detu} \xi(\chi) = \hat{\rm Tr} \left[\hat{U}_\chi
\left(\beta,0\right)\right] = \det \left[{\bf 1}+{\bf
U_\chi}\left(\beta,0\right)\right].
\end{equation}
The last equality can be proven by expanding the determinant
\cite{gubernatis}. Samples are taken by a random walker that
travels through $\chi$-space, taking a new value $\chi_{\rm new}$
from the previous one $\chi_{\rm old}$ if the ratio
\begin{equation}
\label{ratio} r = \frac{W(\chi_{\rm new})}{W(\chi_{\rm old})}
\end{equation}
is larger than one, or else, if $r<1$, taking on $\chi_{\rm new}$
with probability $r$. It can be shown \cite{koonin} that the
sequence of values the walker takes is distributed according to
the weight function $W\left(\chi\right)$, which is typically
chosen to be as close to the integrand as possible to increase the
efficiency of the procedure. Since the consecutive steps are
correlated, the walker has to travel several steps before another
sample is taken to de-correlate them. The average of an observable
(\ref{therobs}) is then simply
\begin{equation}
\label{mcsum}
 \langle \hat{O}\rangle = \frac{\sum_{i} \langle
 \hat{\cal O} \rangle_{i} \Phi_{i}}{\sum_{i}
 \Phi_{i}}
\end{equation}
in terms of its $i$th Monte Carlo sample $\langle
 \hat{\cal O} \rangle_{i}$ and
\begin{equation}
\Phi =  \left\{ \begin{array}{ll}  \frac{G(\chi)
\xi(\chi)}{W\left(\chi\right)} & {\rm for \; continuous \; HS}\\
\frac{\xi(\chi)}{W\left(\chi\right)} & {\rm for \; discrete \;
HS}.
\end{array}
\right.
\end{equation}
Note that $\Phi$, which is just the sign of the weight $W$, can
generally be negative or even complex.

The numerical determination of Eq.~(\ref{mcsum}), which is the
Monte Carlo equivalent of the integrals in Eq.~(\ref{therobs}),
can be difficult in certain situations, even with Monte Carlo
methods: If $S_\alpha = \pm i$ (which generally corresponds to a
repulsive on-site and an attractive next-neighbor interaction, cf.
Eq.~(\ref{salpha})), propagators for the potential
(Eq.~(\ref{uvc0})) contribute negative or complex elements to
$\hat{U}_\chi \left(\beta,0\right)$ (see Eq.~(\ref{detu})). The
integrands in both numerator and denominator are oscillatory, and
the integrals can add up to small numbers. A numerical evaluation
with Monte Carlo methods causes large uncertainties because the
methods are of a stochastic nature, and the number of samples in a
computation remains finite. This is a complication associated with
these methods when the Hubbard-Stratonovitch transformation is
used. It is referred to as the Monte Carlo sign problem. A
pragmatic solution has been used for the shell model Monte Carlo
method to handle this complication \cite{physrep}.

There has been significant effort in stabilizing and optimizing
the Metropolis algorithm for lattice calculations, as they have
been heavily used for models of interacting electrons in condensed
matter physics. Many of the techniques have directly been applied
to NMMC, because the models are similar. Besides using the
checkerboard breakup \cite{gubernatis} technique for kinetic and
spin-exchange parts, we use the Green's function algorithm
described in \cite{srwhite} to reduce the computational burden.

\section{Numerical Results}
\label{nmresult} We now show that for symmetric nuclear matter
(SNM) the energy per particle can be reproduced quite well over a
wide range of densities, and the energy for pure neutron matter
(PNM) for the same potential is discussed. Then, several
observations are presented that give evidence of a first-order
phase transition from a Fermi gas to a clustered system at a
critical temperature $T_c \sim 15 \; {\rm MeV}$. Furthermore, the
symmetry energy and first sound are discussed.

The extensive search in the space of potential parameters included
all components of the central part and spin-exchange. The effort
focused on reproducing saturation density and energy correctly.
The project is considered to be a first step towards a realistic
calculation as indicated earlier. A more realistic calculation has
to contain more parts of the nuclear potential and the spatial
resolution has to improve; a perfect fit over a wide range of
densities should not be expected at this point. The fit has been
performed in such a manner that the saturation energy and density
is reproduced, and that the overall energy curve has a reasonable
form: for sub-saturation densities, the matter should be unstable,
while for $\rho > \rho_0$ SNM should evolve in an unbound state
($E/A > 0 \; {\rm MeV}$ for $\rho \geq 0.4 \; {\rm fm^{-3}}$).

The sign problem unfortunately forces the use of a nuclear
potential that might contradict the usual physical understanding
and intuition based on few-nucleon potential models. It is
generally known that the central potential has a strong repulsion
for short distances and features a long-range attraction. Here,
the desire to avoid the sign problem generates the opposite:
on-site attraction and next-neighbor repulsion. On the other hand,
an on-site attraction and next-neighbor repulsion may not be
unreasonable given the fact that there have been several
mean-field calculations of nuclear matter with the Skyrme forces.
Skyrme forces simulate the interaction with a $\delta$-like
attraction and a $\nabla^2 \delta$-like repulsion. In the lattice
discretization of this investigation, the position of the nucleons
at the same site is only determined up to a cube of size $a$.
Therefore, the on-site potential parameter can be seen as an
average potential within that cube, and by tuning the lattice
spacing accordingly, it could be possible that this parameter is
negative. The exact definition of the parameter depends on a
regularization scheme. In such a scheme the Schr\"odinger equation
has to be solved on a lattice, and by identifying scattering
amplitudes, one could determine the potential parameters from
scattering lengths and effective range \cite{mueller2}. With
respect to the lattice spacing $a$ two approaches can be taken:
First, we constrain ourselves to a description with a fixed number
of lattice sites. Then $a$ becomes a free fitting parameter like
the potential parameters, and the latter would have to be
interpreted as an average potential, as noted above. In the second
approach, $a$ is an discretization parameter for the potential,
and the ultimate goal would be to increase the number of lattice
points with decreasing $a$, getting a smooth parameterization of
the potential. If, in that case, a positive on-site parameter is
used, emulating a hard core repulsion, one has to deal with
oscillatory integrands and commensurately large error bars.

The following potential parameters were obtained:
\begin{eqnarray}
V_c^{(0)} &=& -181.5 \; {\rm MeV \,fm^3},\\ V_c^{(2)} &=& 37.8 \;
{\rm MeV \, fm^5},\\ V_\sigma^{(0)} &=& -31.25 \;{\rm MeV \,
fm^3},\\ V_\sigma^{(2)} &=& 0.0 \; {\rm MeV \, fm^5}.
\end{eqnarray}
All calculations were done with this set of parameters. The
lattice has a spacing of
\begin{equation}
a = 1.842 \; {\rm fm},
\end{equation}
tuned such that quarter filling of the lattice is at saturation
density $\rho_0 = 0.16 \; {\rm fm^{-3}}$. In this paper, the
lattice spacing is an additional fitting parameter, and several
other settings have been tested. However, quarter- and
half-filling at saturation density have a special significance
because certain lattice occupations of the nucleons result in an
energetically favored configuration.

Because of limited CPU time, the calculation is restricted to
$4\times4\times4$ lattices for the moment. This lattice comprises
$10^{38}$ many-body states, and 11520 auxiliary fields are used
for this set of parameters. All calculations are prepared by a
pre-thermalization of the system for 100 steps before we took
measurement samples. Between measurement samples, 15
de-correlation steps have been taken to guarantee statistical
independence of the samples. The autocorrelation of $k$
consecutive samples \cite{koonin}
\begin{equation}
C_{\cal{O}}\left(k\right) = \frac{\langle {\cal{O}}_i
{\cal{O}}_{i+k} \rangle - \langle {\cal{O}}_i \rangle^2}{\langle
{\cal{O}}_i^2 \rangle - \langle {\cal{O}}_i\rangle^2},
\end{equation}
with $i$ being the summation index over samples, has been
monitored for all observables $\cal{O}$ and was held below $10
\%$.

We are also restricted by the fact that the Monte Carlo
simulations cannot be extended to arbitrarily low temperatures.
Even though the numerical routines are quite stable, it is not
possible to add an arbitrary number of time slices, since it
involves more and more matrix multiplications which become
increasingly numerically unstable. In the present case we take
$\Delta \beta = 0.01 \; {\rm MeV^{-1}}$ and were able to go down
to a value of $\beta = 30\times \Delta \beta = 0.3 \;{\rm
MeV^{-1}}$ without running into numerical instabilities. Thus, the
temperature range of the investigation is
\begin{equation}
3.0\; {\rm MeV} \leq T \leq 100\; {\rm MeV}.
\end{equation}

\begin{figure}
\centerline{\psfig{file=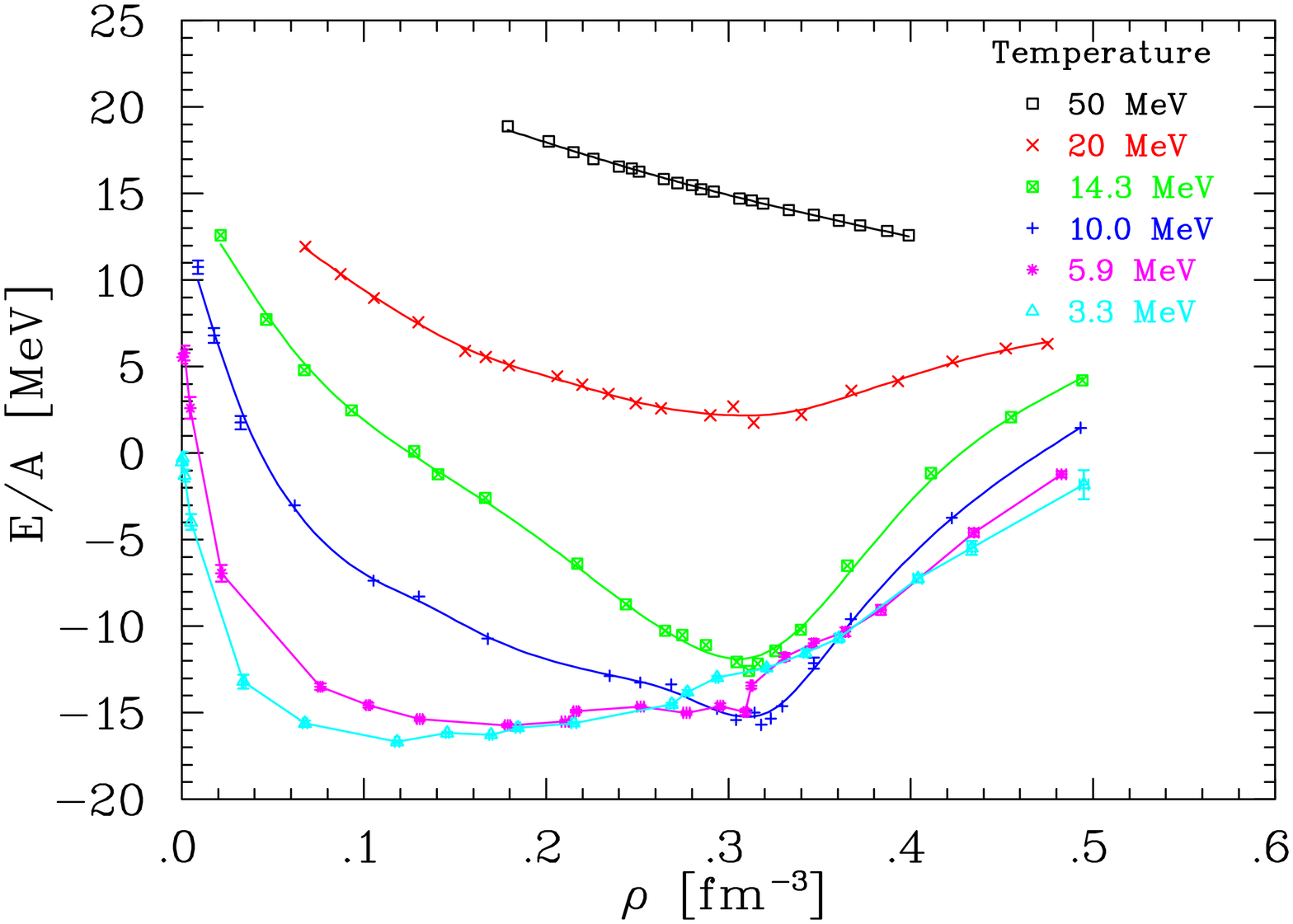,width=7.0in,angle=90}}

\vspace{0.5cm}

\caption[Energy per particle $E/A$ for symmetric nuclear
matter.]{$E/A$ for symmetric nuclear matter as a function of
density $\rho$ and for different temperatures. The purpose of the
lines is to guide the eye.} \label{snmenergy}
\end{figure}

Fig.~\ref{snmenergy} shows the best fit we obtained. With
decreasing temperature, the system develops a minimum at $\rho =
0.32\; {\rm fm^{-3}}$ first, which is most pronounced between
$10-14 \; {\rm MeV}$, before it shifts to lower densities. At
$T=3.3 \; {\rm MeV}$ and $T=5.9 \; {\rm MeV}$ the minimum is very
broad, making matter softer (see also compressibility,
Fig.~\ref{snmcompress} below). For high temperatures and/or high
density, the simulation suffers from the fact that it runs out of
model space: At $T=50\; {\rm MeV}$ the system behaves almost like
a Fermi gas and the energy per particle should behave like $\sim
\rho^{2/3}$. Yet, the curve bends down. Also, for all other
temperatures, the curves converges to the energy of the full
lattice state, $E/A = 5.96 \; {\rm MeV}$, as density increases.
For sub-saturation densities the model gives more binding if
compared to other calculations (see, for example, Refs. \cite{wff}
and \cite{aakmal}), and the energy is not as high for densities
beyond saturation. At $\rho = 0.32 \; {\rm fm^{-3}}$, $E/A$ as a
function of temperature has a minimum at $T \approx 10 \; {\rm
MeV}$ which means that at even lower temperatures $E/A$ increases
again. This contradicts intuition because it would mean that the
system is in an unphysical state.

\begin{figure}
\centerline{\psfig{file=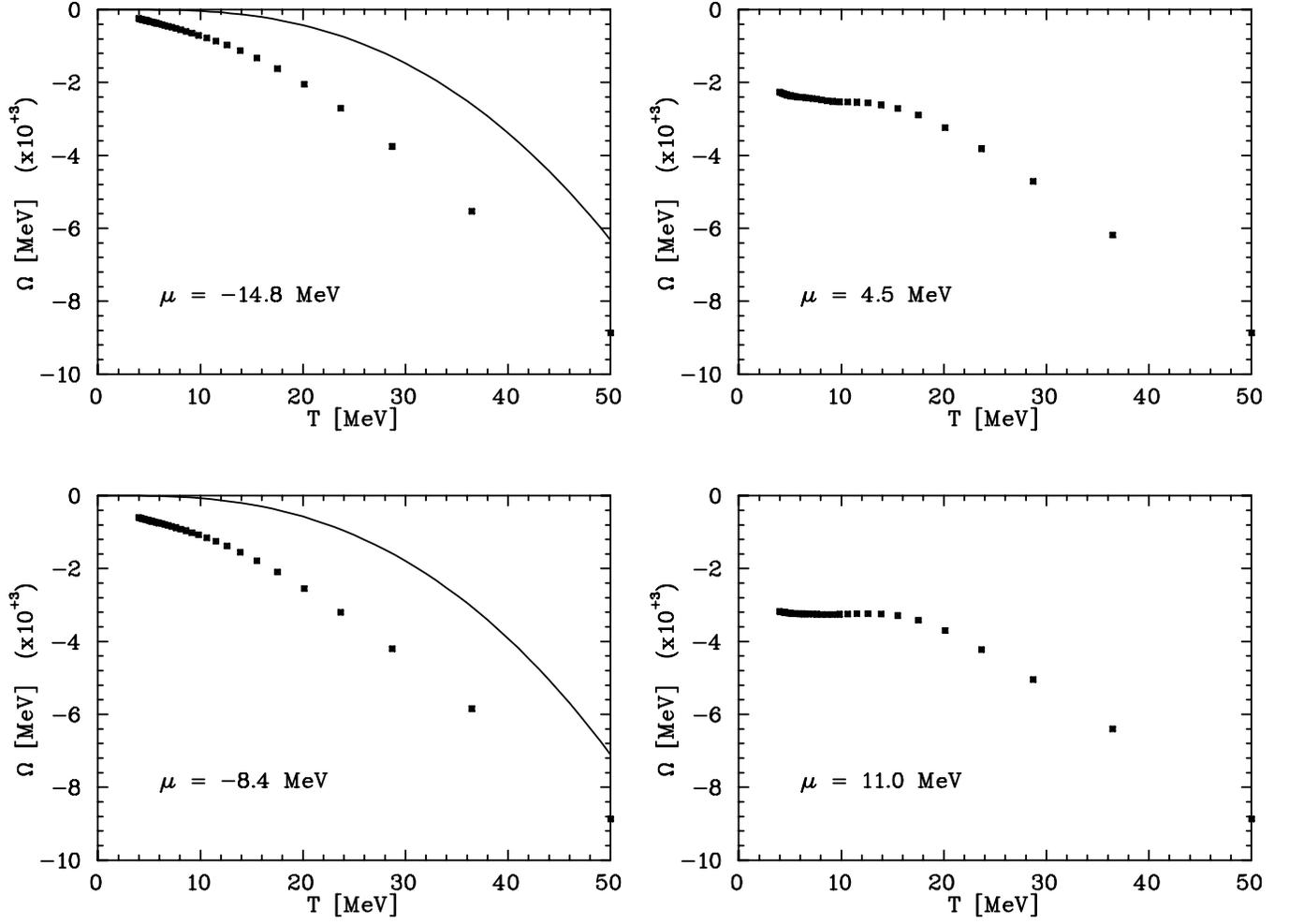,width=7.0in,angle=90}}

\vspace{0.5cm}

\caption[Grand canonical potential for symmetric nuclear
matter.]{Grand canonical potential of symmetric nuclear matter for
different chemical potentials $\mu$. The solid lines represent the
potential for a noninteracting Fermi gas in continuous space.}
\label{snmgp}
\end{figure}

\begin{figure}
\centerline{\psfig{file=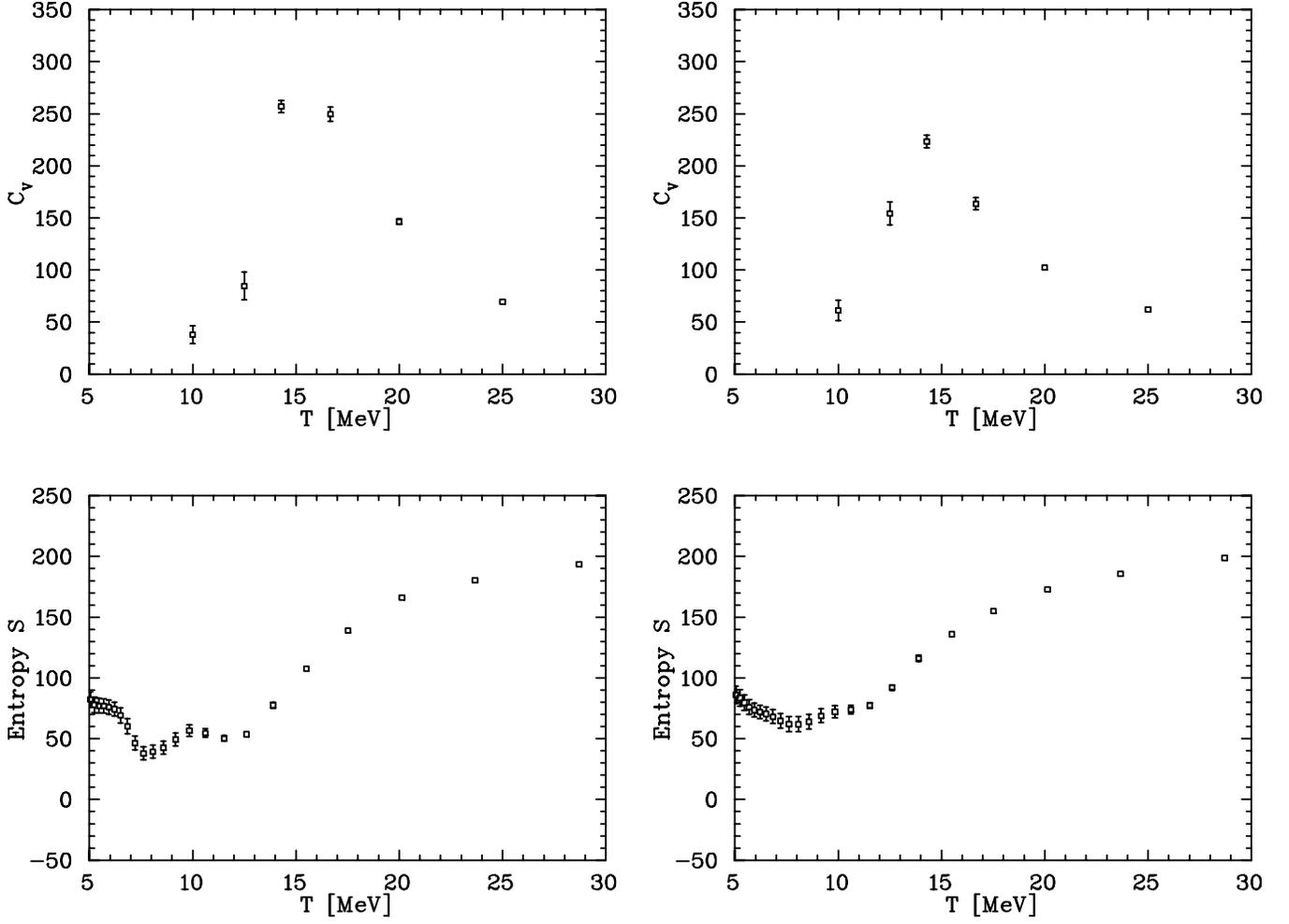,width=7.0in,angle=90}}

\vspace{0.5cm}

\caption[Heat capacity and entropy for a finite piece of
symmetrical nuclear matter.]{Heat capacity and entropy for a
finite piece of symmetrical nuclear matter. The two graphs on the
left show the case $\mu = 0.0 \; {\rm MeV}$, the right ones for
$\mu = 4.0 \; {\rm MeV}$. The heat capacity (upper panels) shows a
distinct maximum, the entropy (lower panels) a relatively sharp
drop with decreasing temperatures.} \label{snmcvandent}
\end{figure}

The last issue needs further explanation. The energy per particle
is not the correct quantity in order to address the question of
stability. Particles fluctuate in and out of the system
differently at different temperature, leaving the average number
of particles unchanged, but contributing to the two-body part of
the Hamiltonian. If, however, the grand potential is plotted (see
Fig.~\ref{snmgp}),
\begin{equation}
\Omega\left(\beta, \mu\right) = -T \ln {\cal{Z}}\left(\beta,
\mu\right),
\end{equation}
with
\begin{equation}
\ln {\cal{Z}} \left(\beta, \mu\right) - \ln {\cal{Z}} \left(0,
\mu\right) = - \int_0^\beta {\rm d} \beta^\prime
E\left(\beta^\prime, \mu\right),
\end{equation}
it turns out to actually be a monotonic function of temperature,
with a slight deviation at $\mu = 11.0 \; {\rm MeV}$ where the
negative slope of $\Omega$, the entropy
\begin{equation}
  S = -\left(\frac{\partial \Omega}{\partial T}\right)_{\mu, V},
\end{equation}
becomes zero between $10 \; {\rm MeV}$ and $14 \; {\rm MeV}$ and
positive again for even lower temperatures. This is a slight
anomaly (see also Fig.~\ref{snmcvandent}) which may have been
caused by the onset of numerical instabilities at low temperatures
or the fact that the lattice spacing is so big and the number of
sites so small that the discretization of space is not accurate
enough.

\begin{figure}
\centerline{\psfig{file=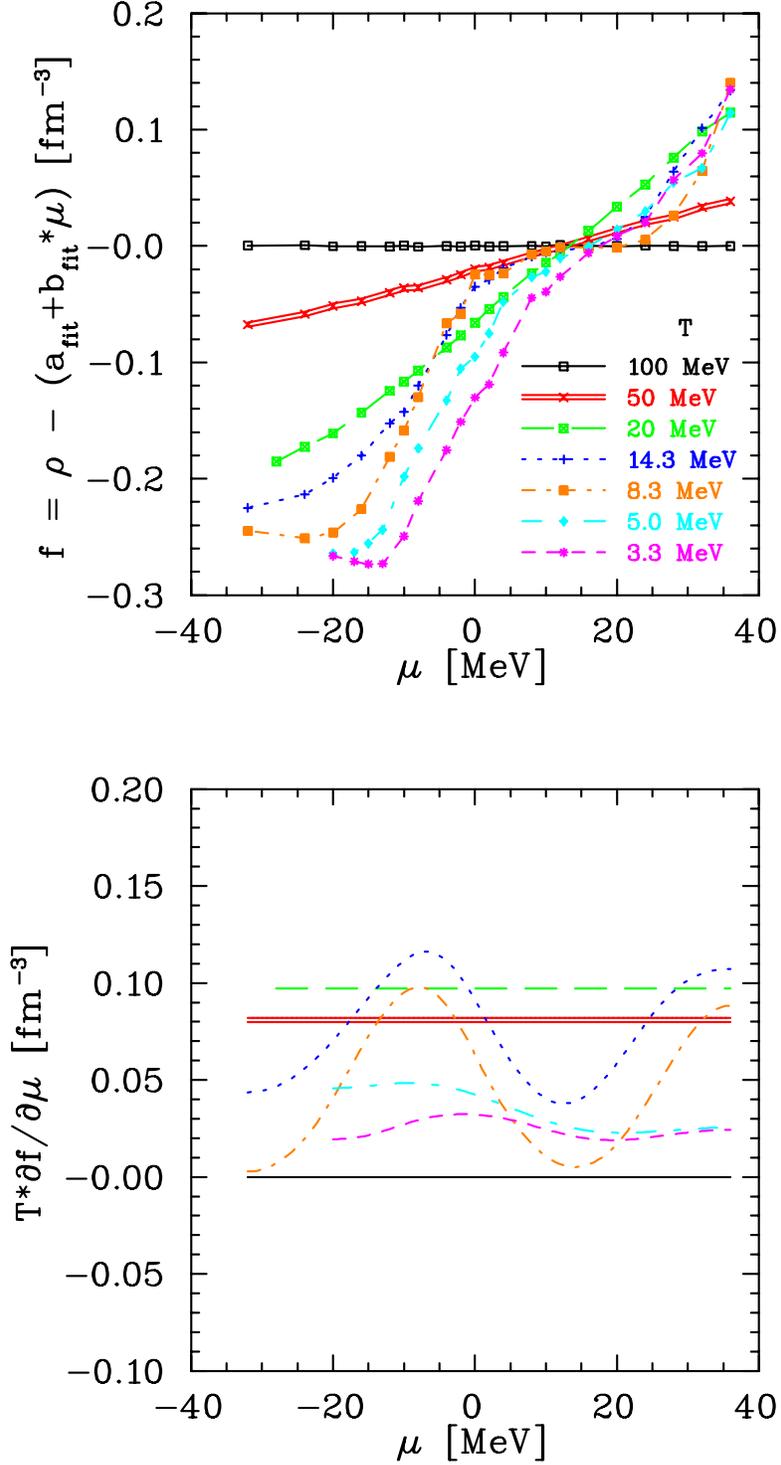,width=4.0in}}

\vspace{0.5cm}

\caption[Density fluctuations in symmetric nuclear
matter.]{Density fluctuations in symmetric nuclear matter. The
upper panel displays the modified density $f$ while the lower
panel shows the derivatives of $f$ with respect to chemical
potential $\mu$ which are proportional to the fluctuations.}
\label{snmdenssig}
\end{figure}

We now introduce several observations which indicate that the
system may undergo a first-order phase transition towards a
clustered system when the temperature is lowered. First, we
investigate changes in density with respect to the chemical
potential $\mu$. It is well known that they are proportional to
particle fluctuations
\begin{equation}
\sigma^2_{N} = T \left. \frac{\partial \langle N \rangle}{\partial
\mu} \right|_{T,V} \sim \left. T \frac{\partial \rho}{\partial
\mu} \right|_{T,V}.
\end{equation}
Such fluctuations are typical for first-order phase transitions
and indicate that particles move between the two phases without
energy cost. For an infinite system, the fluctuations should
diverge, but not for a finite system. In the present case, we
expect particles building clusters and breaking them up again, so
one phase --- the gas phase --- would be that of independent
particles, the other one that of clusters. Since we observe the
single particle density, $\sigma^2_{N}$ describes the fluctuations
in the gas phase in which $\rho$ is linear in $\mu$. At $T=100 \;
{\rm MeV}$, we are in the gas phase with nucleons behaving like a
Fermi gas. Therefore, to simplify the graphs, we have first fitted
the data at $T=100 \; {\rm MeV}$ to a linear function,
\begin{equation}
\rho_{\rm fit} = a_{\scriptscriptstyle \rm fit} +
b_{\scriptscriptstyle \rm fit} \times \mu,
\end{equation}
and then subtracted this function from all data points of all
temperatures, defining a function of temperature and chemical
potential
\begin{equation}
f\left(T, \mu\right) = \rho\left(T, \mu\right) - \rho_{\rm fit}.
\end{equation}
The upper panel of Fig.~\ref{snmdenssig} shows the outcome of this
procedure. We then take the derivative of $f\left(T, \mu\right)$
with respect to $\mu$ and multiply with $T$, and this is shown in
the lower panel of Fig.~\ref{snmdenssig}. The fluctuations show a
pronounced maximum for $T=14.3 \; {\rm MeV}$ and $\mu \approx -8
\; {\rm MeV}$, while they are low for $T= 3.3 \; {\rm MeV}$ and
$T=100 \; {\rm MeV}$. The phase transition seems to occur
somewhere between $T = 8 \; {\rm MeV}$ and $T=20 \; {\rm MeV}$.

Another quantity that suggests the existence of a transition is
the compressibility which is given by
\begin{equation}
\kappa = 9 \rho^2 \left. \frac{\partial^2 E/A}{\partial
\rho^2}\right|_{\rho=\rho_{\rm sat}},
\end{equation}
where $\rho_{\rm sat}$ is the saturation density. We have fitted
the minima of each energy curve to a quadratic function
\begin{equation}
\left. \frac{E}{A} \right|_{\rm fit}\left(\rho\right) = a + b
\times \left(\rho-\rho_{\rm sat}\right)^2,
\end{equation}
and determined the compressibility as $\kappa = 18\rho_{\rm
sat}^2\times b$. All data points in Fig.~\ref{snmcompress} were
obtained with a $\chi^2$ per degree of freedom of less than $1.5$.
Again, a maximum in compressibility (which is in fact an
incompressibility) is observed at $T\approx 14 \; {\rm MeV}$: The
clusters that form repel each other through the next-neighbor
interaction which is repulsive. At $T< 14 \; {\rm MeV}$, matter
becomes softer again due to a broadening of the minima in $E/A$.
This can be explained if one assumes that the system becomes more
dilute. Note that the values of $\rho_{\rm sat}$ change with
temperature.

Finally, we present the heat capacity and entropy of the system.
For a first-order phase transition, the continuous and infinite
system shows a divergence in the heat capacity
\begin{equation}
c_V = \left. \frac{\partial E}{\partial T}\right|_V,
\end{equation}
and a discontinuity for the entropy (with an infinite derivative
at $T_c$),
\begin{equation}
S\left(\beta, \mu\right)= \ln {\cal{Z}} + \beta\langle
H\left(\beta, \mu\right)\rangle - \beta\mu \langle N\left(\beta,
\mu\right)\rangle.
\end{equation}
For a finite system only a maximum in the heat capacity is
expected, and a relatively sharp drop in entropy with decreasing
temperature. Both facts can be verified in Fig.~\ref{snmcvandent}:
The heat capacity suggests a critical temperature of $T_c = 15 \;
{\rm MeV}$, as does the entropy. For the graphs of entropy one has
to keep in mind that the system investigated is quite small (it is
a $4\times4\times4$ lattice only), but two levels, $S\approx 75$
from $T = 5 \; {\rm MeV}$ to $T= 12 \; {\rm MeV}$ and $S=175$ from
$T = 20 \; {\rm MeV}$ to $T= 30 \; {\rm MeV}$ with a steep
decrease in between, can definitely be identified. To show that
this is indeed a first-order phase transition, the calculation has
to be repeated for a larger number of lattice sites, and then it
has to be demonstrated that the drop between the two levels
becomes steeper and steeper, finally resulting in a step-like
function. This will have to be left for a future project. Studies
of a lattice gas model \cite{gulminelli} have shown that finite
size effects do induce anomalies in physical quantities, and a
first-order phase transition rather appears as a second-order one.
The anomaly below $T = 10 \; {\rm MeV}$ has been addressed when
discussing the grand potential. However, the latter quantity shows
a qualitative behaviour at $\mu = 11 \;{\rm MeV}$ that is expected
for a phase transition (cf Fig.~\ref{snmgp}). The infinite system
has a kink (the derivative is not continuous) in the grand
potential at the critical temperature. Consequently, all
quantities consistently suggest a phase transition at a critical
temperature of $T_c \approx 15 \; {\rm MeV}$.

\begin{figure}
\centerline{\psfig{file=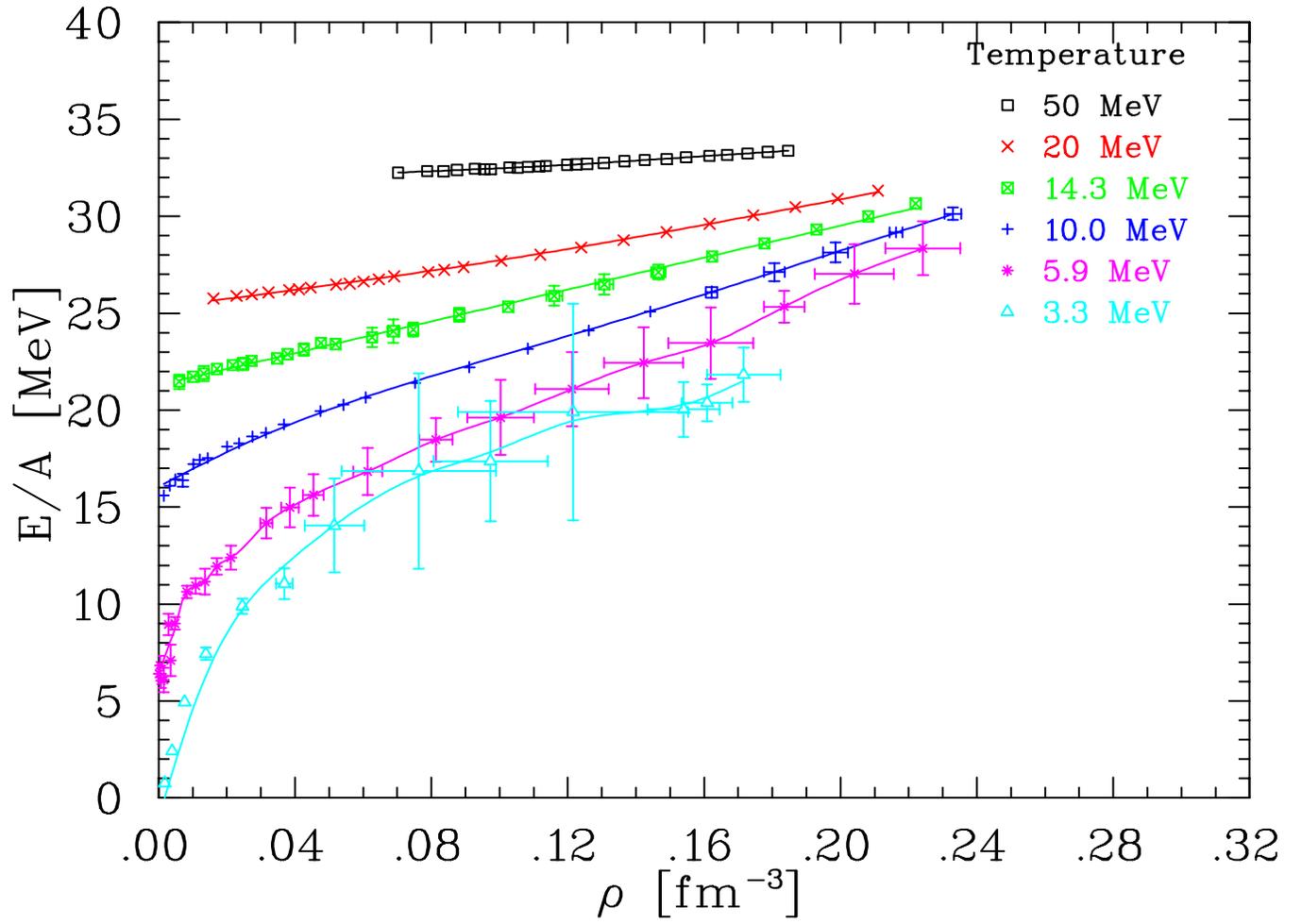,width=7.0in,angle=90}}

\vspace{0.5cm}

\caption[Energy per particle $E/A$ for pure neutron matter.]{$E/A$
for pure neutron matter as a function of density $\rho$ and for
different temperatures. The lines guide the eye.}
\label{pnmenergy}
\end{figure}

In Fig.~\ref{pnmenergy} we show the energy per particle for pure
neutron matter. The uncertainties for this case are much larger
than for symmetrical nuclear matter. As a potential, we used the
parameters obtained from the fit to symmetric nuclear matter, even
though we could have fitted the potential parameters for this case
anew, including an isospin-exchange potential. Therefore we view
the results for pure neutron matter more as a test to see how well
the given potential already reproduces the energy. Note that the
slopes of the curves at high temperatures are not negative as it
is for symmetrical nuclear matter. But clearly, we cannot conclude
that the energies at $T= 3.3 \; {\rm MeV}$ have converged to that
of the ground state because the curve differs quite a bit from
that of $T = 5.9 \; {\rm MeV}$. At the lowest temperature they are
$4-5 \; {\rm MeV}$ higher than those of the ground state as
calculated in Ref.~\cite{wff}, but the general shape of the curve
is very similar. This is no surprise, since pure neutron matter is
like a Fermi gas, with attractive forces between neutrons lowering
the energies with respect to the non-interacting system. The
search for any kind of phase transition in the range of $5-50 \;
{\rm MeV}$ was to no avail. It is likely that a phase transition
occurs at lower temperature.

\begin{figure}
\centerline{\psfig{file=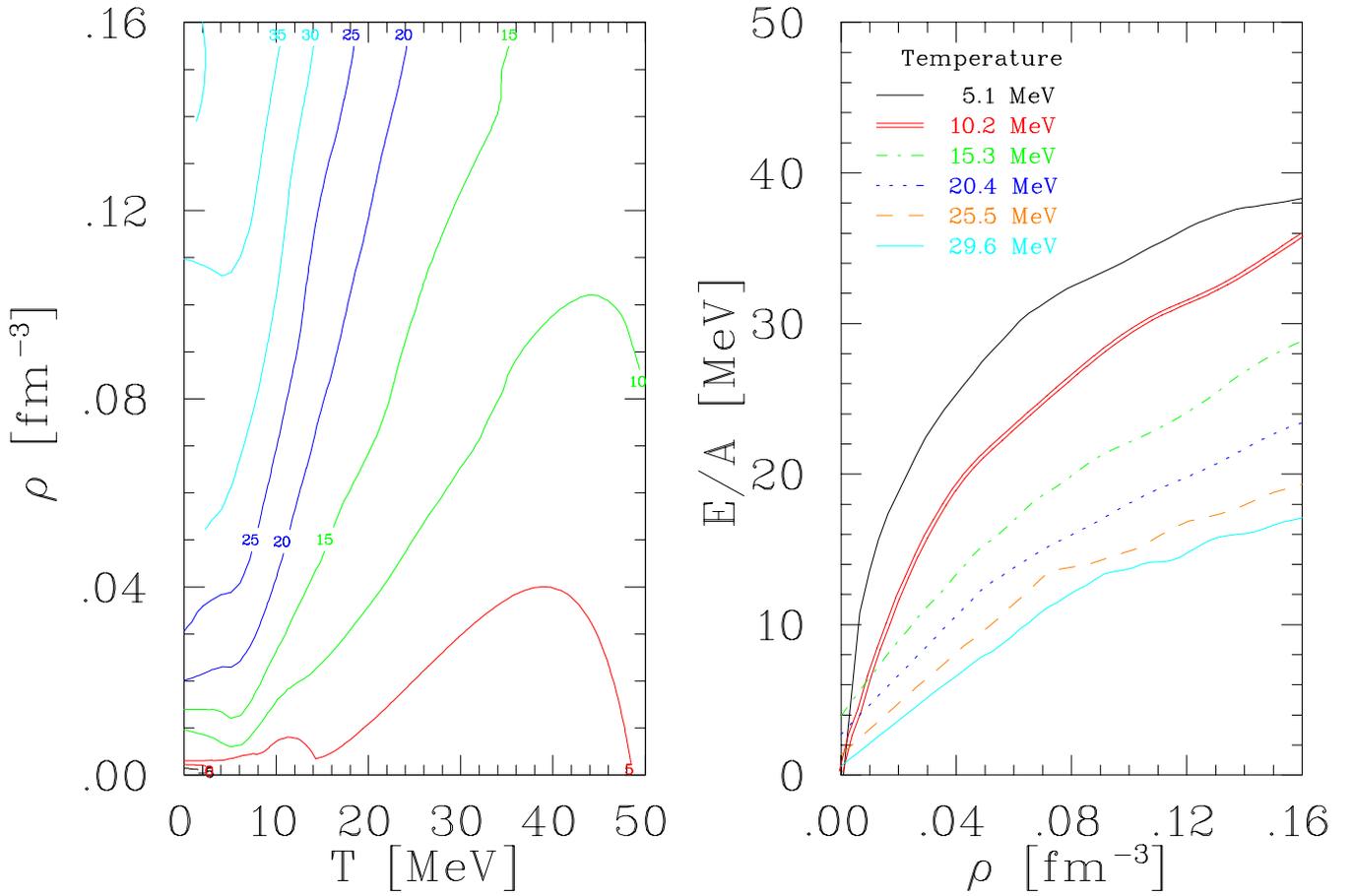,width=7.0in,angle=90}}

\vspace{0.5cm}

\caption[Symmetry energy for symmetric nuclear matter.]{Symmetry
energy for symmetric nuclear matter as a function of density and
temperature. Shown is the coefficient $a_{sym}$ of the
semi-empirical mass formula. The left panel shows a contour plot,
the right one shows one-dimensional cross-sections of it at
different temperatures. $a_{sym}$ is increasing with density and
decreasing with temperature.} \label{snmsymenergy}
\end{figure}

We finally calculate two additional observables and compare them
with other calculations found in the literature. The symmetry
energy, which appears as a coefficient $a_{sym}$ in the
semi-empirical mass formula
\begin{equation}
\frac{E_{sym}}{A} = a_{sym}\frac{\left(N-Z\right)^2}{A^2},
\end{equation}
is plotted in Fig.~\ref{snmsymenergy}. We used the energy per
particle of pure neutron matter (PNM) and SNM, subtracted them and
interpolated the result on a mesh. Since the error bars for pure
neutron matter are larger for low temperatures, the graphs should
be viewed with caution. Nevertheless it appears that the symmetry
energy is increasing with density and decreasing with temperature,
as one would expect. Indeed, at high temperature SNM and PNM both
are more like a Fermi gas, and only at low temperatures do they
become different. The observed dependence on density can be
explained by the fact that a dilute system is barely interacting
while the probability of clustering increases with density. At
saturation density and low temperature, we obtain a coefficient of
\begin{equation}
a_{sym} \approx 38 \pm 3 \; {\rm MeV},
\end{equation}
which is not too different from the generally accepted value
\cite{schuck2} of $a_{sym} = 28.1\; {\rm MeV}$. This discrepancy
is in part due to the fact that the calculations for pure neutron
matter have not converged completely.

The first sound velocity has been calculated using the formalism
of relativistic fluid dynamics:
\begin{equation}
u/c = \sqrt{\left. \frac{\partial p}{\partial e} \right|_S},
\end{equation}
where
\begin{equation}
e = \rho \times \left(m_Nc^2 + \frac{E}{A}\right)
\end{equation}
and
\begin{equation}
p = \rho^2 \left. \frac{\partial E/A}{\partial \rho} \right|_S.
\end{equation}
In general, the first sound we obtain is too low compared to Ref.
\cite{wff,eosnes}, but the velocities are of the same order of
magnitude. Several calculations \cite{wff,eosnes} show a violation
of causality at a few multiples of the saturation density, and so
do ours. Our results (Fig.~\ref{snmsound}) show the correct
temperature dependence in the sense that it conforms with the
compressibility: higher sound speed for intermediate temperatures
(high incompressibility) and lower speeds for both low and high
$T$.

\section{Discussion and Conclusion}
The model considered in this paper is an exact treatment taking
first steps towards a more realistic Hamiltonian, and it is an
improvement compared to previous calculations. Nevertheless it can
be extended to include more physics, and details in the algorithm
can be improved. First of all, more computer power is necessary to
reduce finite size effects. A lattice of $10\times10\times10$
points would be desirable, and also the imaginary time dimension
could be pushed further. This requires stable matrix techniques.
The present code can handle 30 time slices comfortably using
commonly known sparse matrix techniques. But, as the lattice
spacing decreases, one needs to go to larger imaginary time to
separate the ground state from excited states. At the same time it
is not possible to increase $\Delta \beta$ as it would induce
finite time effects. Therefore, an improved effective matrix
algorithm would be needed to allow for more matrices to be
multiplied. Along with a bigger lattice, the resolution of the
potential could be increased, resulting in
next-to-nearest-neighbor and further interactions. This extension
of the spatial dependence of the potential can easily be
accomplished and is only restricted by computational power.

Another big hurdle is the sign problem. The solution to this
obstacle will result in a huge advancement in many areas of
computational physics and chemistry. Rom et al. \cite{rom} have
made some progress which could prove beneficial for the model
described here too: It basically consists of shifting the contour
of the auxiliary field integrals, which is equivalent to
subtracting a mean-field from the problem. We plan to investigate
this method and its application to nuclear matter more rigorously
in the future.

The physics of nuclear matter itself is certainly more involved
than the current model can account for. Mesons are not included as
explicit degree of freedom, and the various exchanges are only
simulated indirectly through the choice of potential and its
parameters, very much like in AV18 or other potentials. Realizing
that the auxiliary fields behave like massless bosons, one could
ponder how a Monte Carlo procedure would look like that includes
meson exchange directly. Such a procedure could be quite similar
to already established auxiliary field Monte Carlo procedures.

It is known that three-body and perhaps higher-order many-body
forces are important to describe saturation properties of nuclear
matter correctly. However, incorporating these forces in a Monte
Carlo calculation is currently impossible, basically because there
is no scheme to reduce higher-order forces to the single particle
formalism. Such a scheme could lie in a multiple application of
the Hubbard-Stratonovitch transformation for a single many-body
interaction. But as long as such a scheme is not available, an
approximation could be established on top of this two-body
calculation that incorporates higher-order effects. A first
attempt would be to calculate the three-body contribution to the
energy obtained from the one-body densities of this Monte Carlo
calculation and a given three-body Hamiltonian.

In conclusion, this project has produced promising results which
should be viewed as a starting point to an exact solution of
infinite nuclear matter. In a model with a relatively simple
Hamiltonian, and further limited by a very small lattice, we were
able to reproduce saturation properties of symmetric nuclear
matter. The energy of pure neutron matter, using the same
potential, gave reasonable results, even though it had not yet
converged. Furthermore, we presented evidence in form of
mechanical and thermodynamical observables which support the
existence of a phase transition from a Fermi gas to a clustered
system. Particle fluctuations of the gas phase seem to reach a
maximum at $T \approx 14 \; {\rm MeV}$. The heat capacity and
compressibility also have a maximum at around this temperature.
Entropy and grand potential show a behaviour as it is expected for
a first-order phase transition. Other quantities like symmetry
energy and first sound velocity show reasonable agreement with
other calculations.

\acknowledgements This work was supported in part by the National
Science Foundation, Grants No. PHY97-22428 and PHY94-20470. The
calculations were performed on a HP Exemplar X-class supercomputer
with 256 nodes, operated by the Center for Advanced Computing
Research at the California Institute of Technology.

\begin{figure}
\centerline{\psfig{file=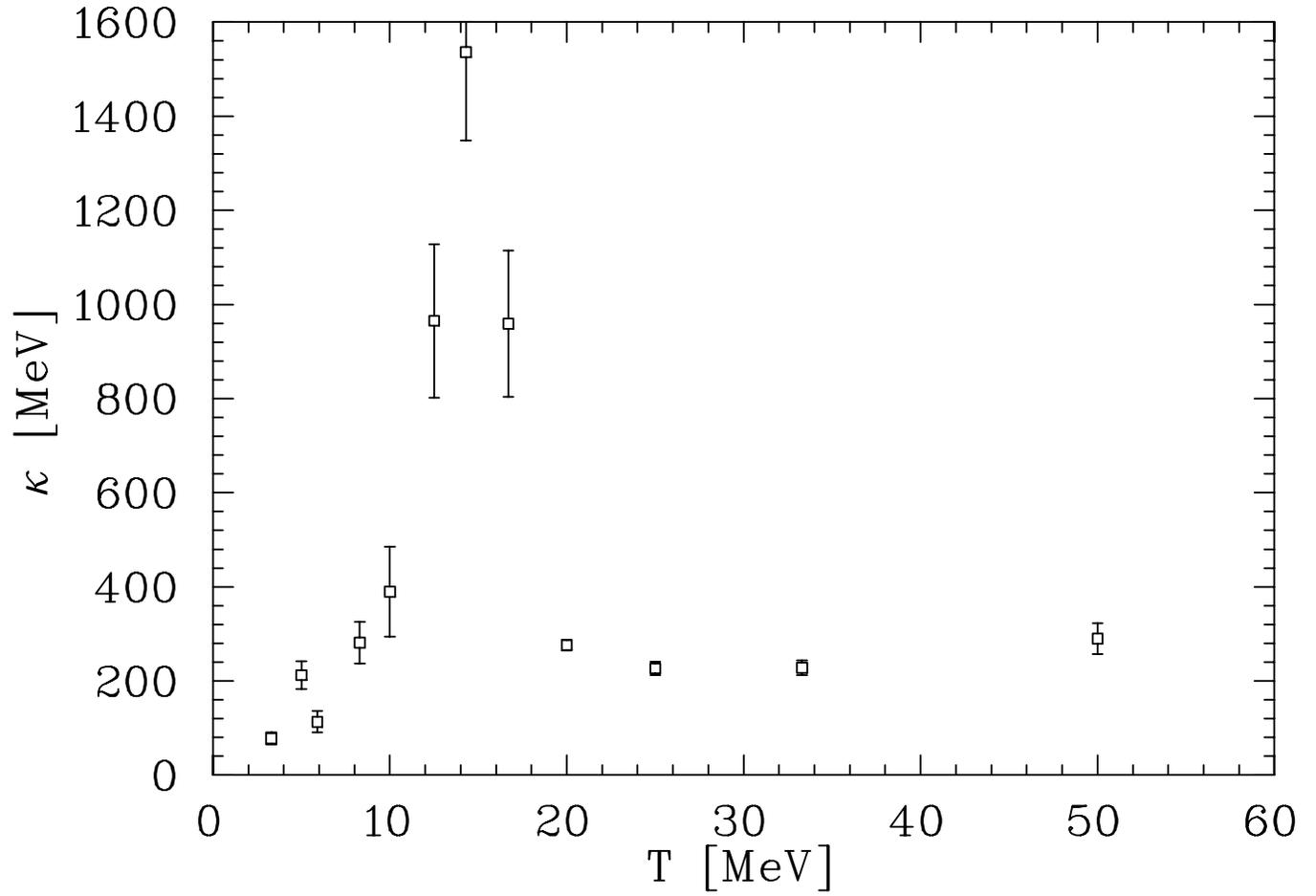,width=7.0in,angle=90}}

\vspace{0.5cm}

\caption[Compressibility of symmetric nuclear
matter.]{Compressibility of symmetric nuclear matter. The minima
of the energy curves have been fit to a parabola with a $\chi^2
\leq 1.5$ per degree of freedom.} \label{snmcompress}
\end{figure}
\begin{figure}
\centerline{\psfig{file=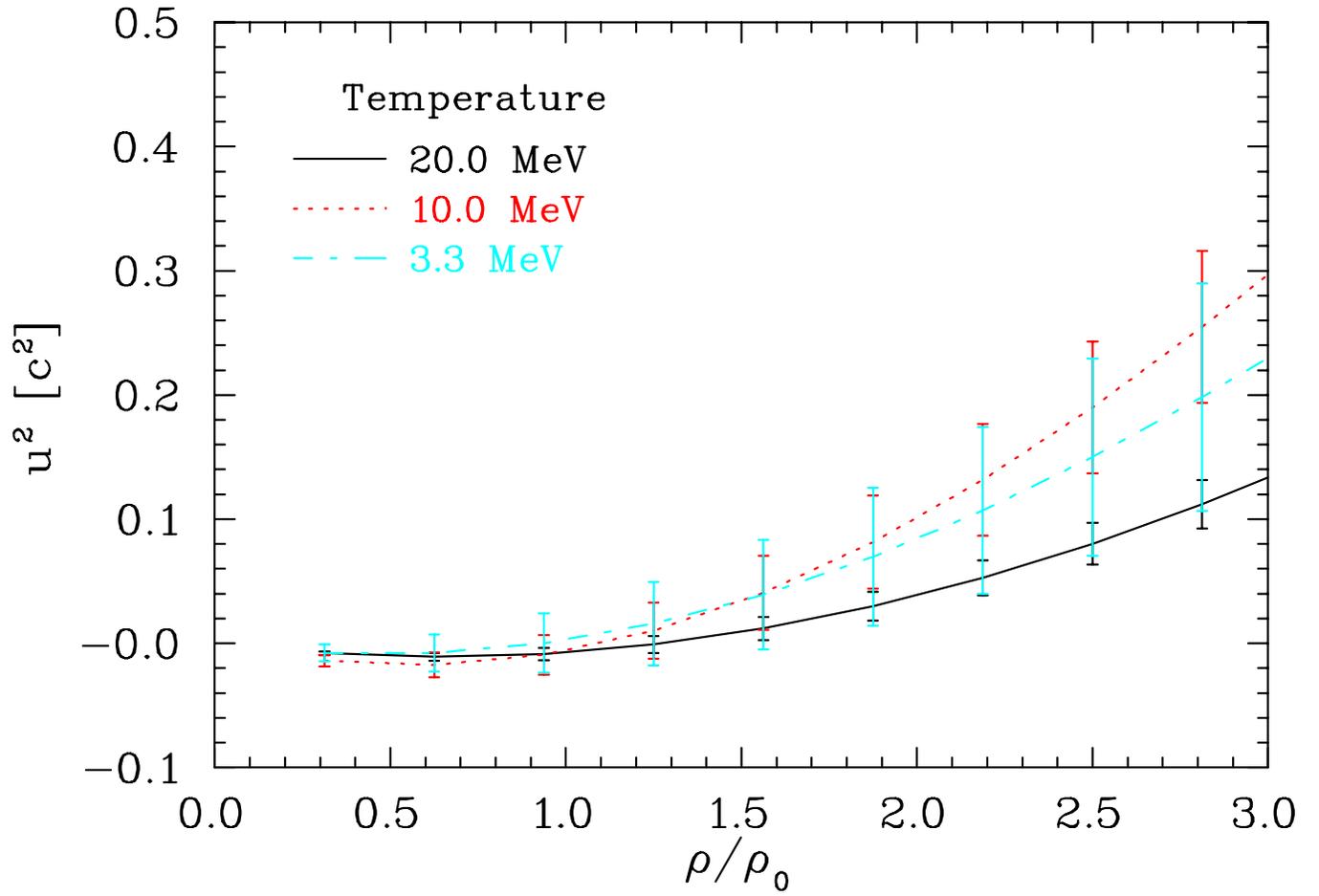,width=7.0in,angle=90}}

\vspace{0.5cm}

\caption[First sound velocity for symmetric nuclear matter.]{First
sound velocity for symmetric nuclear matter. The temperature
dependence of the speed corresponds to the compressibilities as
shown in Fig.~\ref{snmcompress}.} \label{snmsound}
\end{figure}
%
%

\end{document}